\documentclass[%
 reprint,
 amsmath,amssymb,
 aps,nofootinbib,   
]{revtex4-1}
\usepackage{bm}
\PassOptionsToPackage{linktocpage}{hyperref}
\usepackage[hyperindex,breaklinks]{hyperref}
\usepackage{enumitem}
\usepackage{slashed}

\renewcommand{\theta}{\vartheta}

\usepackage{array}
\usepackage{mathtools}

\usepackage{etoolbox}

\begin{document}

\title{On $S$-Matrix Exclusion of de Sitter and Naturalness}

\author{Gia Dvali  
} 
\affiliation{%
Arnold Sommerfeld Center, Ludwig-Maximilians-University,  Munich, Germany, 
}%
 \affiliation{%
Max-Planck-Institute for Physics, Munich, Germany
}%

\date{\today}

\begin{abstract} 

The cosmological constant puzzle, traditionally viewed as a naturalness 
problem, is evidently nullified by the $S$-matrix formulation
of quantum gravity/string theory. 
We point out an implication of this fact for another naturalness 
puzzle, the Hierarchy Problem between the weak and Planck scales.  
  By eliminating the landscape of de Sitter vacua and eternal inflation, the $S$-matrix formulation 
exhibits an obvious tension with the explanations
 based on anthropic selection or cosmological relaxation
 of the Higgs mass. 
 This sharpens the Hierarchy Problem in a profound way.  
 On one hand, it strengthens the case for explanations based on new physics not far from the weak scale. At the same time, it opens up 
 a question, whether instead the hierarchy is imposed by the 
 $S$-matrix consistency between the Standard Model and gravity. 

  \end{abstract}

\maketitle

  In perturbative loop calculations, the vacuum energy-density, 
  $\Lambda$,  is directly  sensitive to the UV-cutoff of the theory.
    This cutoff-sensitivity also makes it un-calculable. Many effects that come from near-cutoff physics, such as, e.g., contributions from virtual micro black holes, can at best be estimated up to unknown numerical factors.  This uncertainty is the source of the Cosmological Constant Puzzle. 
 This puzzle is traditionally viewed as a 
  problem of naturalness. \\ 
 
    Some time ago, a change of the point of view was 
 offered \cite{Dvali:2013eja}. 
  It was suggested that instead the puzzle should be viewed as 
 a matter of consistency of quantum gravity. 
 It was argued that in quantum gravity the de Sitter cannot be 
 viewed as a vacuum but only as  
 an excited (coherent) state constructed on top of a true 
 $S$-matrix vacuum of Minkowski.  \\
 
 It was then found that the validity of semi-classical 
 de Sitter is bounded from above 
 by the anomalous quantum break-time \cite{Dvali:2013eja, Dvali:2017eba}.  This fact severely restricts the potential energy landscape of 
 quantum gravity/string theory. In particular, it excludes
 local and global de Sitter minima, as well as, any scalar potential
supporting an eternal inflation.   The quantitative bounds can be 
found in original papers and are unimportant for the present discussion.  
  The essential message is that a would-be naturalness puzzle 
  turned out to be absent.  \\
  
   This outcome is deeply rooted in the $S$-matrix formulation of 
 quantum gravity, which is organic to string theory.
   The incompatibility of de Sitter with the $S$-matrix formulation of quantum gravity can be understood 
    from a simple scaling argument presented in \cite{Dvali:2020etd}.
   The key point is that the rigidity of de Sitter ``vacuum", as well as,  
   the quantum coupling of gravitons, are controlled by the
   same parameter: The Newton's  gravitational coupling, $G$. \\
   
  A scattering of gravitons, at momentum-transfer $q$, is controlled by 
  the quantity $q^2G$. In order to have a non-trivial $S$-matrix for quantum gravity, this quantity must be non-zero.  This is in conflict 
  with the rigidity limit of de Sitter, 
  \begin{equation} \label{limit} 
   G \rightarrow 0\,, ~  \Lambda \rightarrow \infty\,,  ~
       \Lambda G \equiv  \frac{1}{R^2} = {\rm finite}\,.    
   \end{equation} 
  The last condition indicates that the de Sitter curvature radius, 
  $R$, is kept finite. 
   The equation (\ref{limit}) describes an unique limit in which the back-reaction
on de Sitter state from the scattering process vanishes.  This is a necessary condition for a valid $S$-matrix vacuum.  \\
  
   Obviously, in the limit (\ref{limit}), the quantum gravity trivializes.      
In contrast, there is no {\it a priory}  problem in a short-distance-limit  
$S$-matrix description for the interactions with Wilsonian
UV-completion.  For example,  for 
an asymptotically-free QCD-like theory, with the confinement length larger 
than $R$, such a treatment is valid. \\

  Naturally, the exclusion  of $\Lambda$, no matter how small, from the energy budget 
 of our Universe, sharpens the question of dark energy. 
 Whatever the explanation,  its existence becomes a signal for new physics beyond the framework of the Standard Model +  Einstein gravity.  
 At the same time, the elimination of the simplest explanation in form  
 of $\Lambda$,  should also serve as additional motivation 
 for the high-scrutiny
analysis of cosmic acceleration,  such as \cite{Colin:2018ghy}. \\

  In this note we wish to point out that the above has important implications for another naturalness puzzle, the Hierarchy Problem. 
   Similarly to the cosmological constant puzzle, the 
  Hierarchy Problem also emerges from the quantum sensitivity 
  to the UV-cutoff. In this case, the sensitive quantity is the mass of 
  the Higgs  scalar.  \\
  
   The  naturalness approaches to the Hierarchy Problem can be split in two categories: The ones that require a new stabilizing physics  
   not far from the weak scale and the ones that do not.
   In the second category there are two entries: 
 1) Anthropic selection \cite{Agrawal:1997gf}; and  2) Cosmological relaxation of the Higgs 
 mass \cite{attractor}.  We shall review them separately. \\

The anthropic selection \cite{Carr:1979sg,Vilenkin:1994ua}
  of the Higgs mass was 
suggested by Agrawal, Barr, Donoghue and Seckel
\cite{Agrawal:1997gf}, generalizing the reasoning 
originally applied by Weinberg \cite{Weinberg:1987dv}
 to the puzzle of cosmological constant. 
Various implementations of this idea have been discussed in the 
literature (see, e.g., \cite{ArkaniHamed:2005yv}). 
 We shall not enter in details of model-building,
 but instead  focus  on very general necessary ingredients of the approach. \\
 
The first is the notion that the observed value 
of the Higgs mass (likely in combination with some other parameters),
is crucial for the existence of human observers.
The second ingredient is the mechanism of scanning among 
the different values of 
the Higgs mass.  That is, a theory must allow for a plentitude of 
vacua with different values of the Higgs mass and must also provide a mechanism for actualizing these values in different parts of the Universe. 
The generic mechanism
for this actualization is the eternal inflation \cite{eternalV, eternalL}. \\
 
  In the second class of theories \cite{attractor}, to which we refer to 
as the ``cosmological relaxation of the Higgs mass",  the selection is not anthropic. However, the actualization mechanism is similar.  During a long period of inflation, the values of the parameters are scanned.  \\

The crucial ingredient is that the 
 Higgs mass acts as a control parameter for the scanning process. 
The scanning stops via 
a certain back-propagation mechanism, when the Higgs mass  reaches a desired value (attractor). The driving force of the cosmic evolution 
towards the attractor is the number density of vacua, which diverges in the neighbourhood of the attractor point. 
 In this way, after a sufficiently long period of cosmic inflation, 
 the actualized values of the Higgs mass in different 
inflationary domains,  statistically converge towards the attractor value.  This value is stable
 with respect to quantum corrections, since 
it is determined by non-UV-sensitive 
parameters of the Standard Model, such as the QCD scale.  
 \\ 
 
   We now arrive to our central point. 
 The $S$-matrix formulation of quantum gravity, by excluding  
the eternal inflation and de Sitter vacua, eliminates the possibility 
of an efficient cosmological scanning of the theory parameters. This conclusion applies equally to the scanning of the Higgs mass. 
  Thus, both the anthropic selection and the cosmological 
 relaxation mechanisms become problematic.  
 The least, in both approaches, one needs to rethink the actualization mechanisms of the vacua with different values of the target parameters.\\
  
 In this respect, the attractor mechanism may look somewhat more promising.  Due to an infinite vacuum entropy around the attractor value, one may hope that neither de Sitter vacua nor the process of long cosmic diversification is required. Instead, the Higgs value could be driven 
 towards the attractor domain by a single ``jump" from some initial condition (for example, in the process of creation of an inflationary Universe 
 from ``nothing" \cite{eternalV, Vilenkin:1994ua}).  
  The probability  of such a transition could be enhanced 
 by the divergent entropy factor due to multiplicity of vacua close to 
 the attractor. 
   The above are potentially interesting possibilities, but they require a substantial modification of the scanning mechanism. \\
   
   The bottomline is 
  that the $S$-matrix formulation of gravity turns out to be incompatible with the simplest actualization mechanism based on eternal inflation
 on the landscape of de Sitter vacua. This affects our view of the Hierarchy Problem profoundly.  \\
 
  First,  from the naturalness perspective, the existence of new 
  stabilizing physics, not far from TeV energies, becomes much stronger justified.   
  \\
 
 At the same time, another question opens up.
 Can the hierarchy, instead of being a naturalness issue, 
 be a matter of consistency 
 for embedding the Standard Model in an $S$-matrix theory of 
 gravity? \\
 
  A support for thinking in this direction, is provided by the known 
  cases in which the hierarchy originates  
 from consistency.  For example, in theories with large number 
 of particle species $N$, the Higgs mass is bound not to exceed 
 $1/\sqrt{N}$ fraction of the Planck scale \cite{Dvali:2007hz}.  This bound is imposed 
 by consistency of quantum gravity with well-established features of  black hole physics.  In this light, it is not unimaginable that there exists 
 an independent $S$-matrix reason for the lightness of the 
 Higgs relative to the Planck mass.   
  For example, it is conceivable that a large separation of scales
 is necessary for avoiding a generation of de Sitter minima 
 by quantum gravity effects. \\ 
 
  Whatever the answer, the 
 $S$-matrix nullification of the cosmological constant puzzle, 
 sharpens the question of hierarchies in a non-trivial manner.  \\

 {\bf Acknowledgements} \\
 
 We thank Goran Senjanovi\'c for comments.
This work was supported in part by the Humboldt Foundation under Humboldt Professorship Award, by the Deutsche Forschungsgemeinschaft (DFG, German Research Foundation) under Germany's Excellence Strategy - EXC-2111 - 390814868,
and Germany's Excellence Strategy  under Excellence Cluster Origins.

\end{document}